\documentclass[lettersize,journal]{IEEEtran}
\usepackage{amsmath,amsfonts}
\usepackage{amssymb}
\usepackage{graphicx}
\usepackage{algorithm}
\usepackage{array}
\usepackage[caption=false,font=normalsize]{subfig}
\usepackage{caption2}
\usepackage{textcomp}
\usepackage{url}
\usepackage{verbatim}
\usepackage{cite}
\usepackage{color}
\usepackage{xcolor}
\usepackage[normalem]{ulem} % use normalem to protect \emph
\usepackage{soul}
\usepackage{soul} % 导入 soul 包
\soulregister{\cite}7

\usepackage{booktabs} %三线表
\usepackage{multirow}
\usepackage{tabularx}
\usepackage{nomencl}

\makenomenclature
\renewcommand\nomgroup[1]{\item[\bfseries]}
\begin{document}

\title{A Modular and High-Resolution Time-Frequency Post-Processing Technique}

%%thank放里面后面图片显示不正常
\author{Jinshun Shen and Deyun Wei
\thanks{Jinshun Shen and Deyun Wei are with the School of Mathematics and Statistics, Xidian University, Xi’an 710071, China. }}

%\markboth{IEEE Signal Processing Letters}%


\maketitle

\begin{abstract}
In this letter, based on the variational model, we propose a novel time-frequency post-processing technique to approximate the ideal time-frequency representation. Our method has the advantage of modularity, enabling "plug and play", independent of the performance of specific time-frequency analysis tool. Therefore, it can be easily generalized to the fractional Fourier domain and the linear canonical domain. Additionally, high-resolution is its merit, which depends on the specific instantaneous frequency estimation method. We reveal the relationship between instantaneous frequency estimation and reassignment method. The effectiveness of the proposed method is verified on both synthetic signals and real world signal.
\end{abstract}


%
%\nomenclature[2]{TFR}{TFR}%
%\nomenclature[1]{TF}{TF}%
%\nomenclature[3]{FrST}{Fractional S-transform}%
%\nomenclature[6]{SSSTFT}{Synchrosqueezed STFT}%
%\nomenclature[4]{SSWT}{Synchrosqueezed wavelet transform}%
%\nomenclature[5]{SSST}{Synchrosqueezing S-transform}%
%\nomenclature[7]{SSFrST}{Synchrosqueezing fractional S-transform}%
%\nomenclature[8]{IFE-FrSTSST}{Modulated synchrosqueezing FrST}%
%\nomenclature[9]{${{\rm{FrST}}_\varphi ^\alpha s(t)}(a,b)$}{Transform result of $s(t)$ for the FrST}%
%\printnomenclature

\section{Introduction}
\IEEEPARstart{T}{ime-frequency} representation (TFR) is an effective tool to describe the time-frequency (TF) characteristics of signals \cite{qian1999joint}. Therefore, it is very important to obtain a high TF resolution and clear result. However, traditional methods, such as short-time Fourier transform (STFT) \cite{durak2003short}, continuous wavelet transform (CWT) \cite{daubechies2009wavelet}, S-transform \cite{stockwell1996localization}, chirplet transform \cite{li2022chirplet}, etc., cannot achieve the ideal results.
Therefore, many post-processing tools of time-frequency analysis (TFA) have been developed to approximate the ideal TFR. For example, the reassignment method (RM) \cite{auger2013time} reallocates TF coefficients to the appropriate positions on the two-dimensional (2D) TF plane, for improving TF resolution. Unfortunately, it breaks the ability to reconstruct signals due to rearrangement on 2D plane. Recently, the synchrosqueezing transform (SST) \cite{daubechies2011synchrosqueezed} has been proposed as a special RM to rearrange only in the frequency or scale axis. Thus, SST can recover the signal, leading to its popularity.

Although SST can obtain compact TF results, it suffers from energy diffusion and appears blurred on the TF plane. It is worse for analyzing non-stationary signals. Therefore, many improved methods have been proposed to enhance the TF concentration and obtain high-resolution TFR. In general, there are two prevailing improvements. One option is to use a TFR with better performance to replace STFT or CWT.
For example, adaptive CWT \cite{li2020adaptivec}, adaptive STFT \cite{li2020adaptives}, and adaptive S-transform \cite{wang2018high} are employed instead of the original TFR. 
Another alternative is to carry out the second-order or 
high-order Taylor expansion on the signal to generate a more accurate estimate of the instantaneous frequency (IF), such as the second-order SST \cite{oberlin2015second} and high-order SST \cite{pham2017high} for applications to gravitational-wave signal. Despite the improved TF resolution, the computational complexity is their drawback.

Recently, to achieve the ideal TFR, a novel post-processing method, the synchroextracting transform (SET) \cite{yu2017synchroextracting}, has been proposed, which extracts only the TF coefficients near the IF and discards the rest of TF coefficients. It causes SET to sacrifice the accuracy of reconstructed signal. Subsequently, motivated by SET, the local maximum synchrosqueezing transform (LMSST) \cite{yu2019local} has been presented. Its significant contributions are that a novel reassignment operator is proposed and the IF estimation does not depend on the specific TFR.

In this letter, a modular and high-resolution TF post-processing technique is proposed based on the variational model, which can achieve ideal TFR. The results of our proposed method are not constrained by the performance of TFR and can be easily generalized to the fractional Fourier domain \cite{wei2016generalized,tao2009short} and the linear canonical domain \cite{wei2022linear}, just by replacing the TFA tool. Additionally, we also reveal the relationship between the reassignment operator and IF estimation.

The rest of this letter is organized as follows. In Section II, we recall briefly the theories of STFT and SST. In Section III, the novel TF post-processing technique is introduced. Numerical validation is shown in
Section IV. Finally, the conclusion is shown in Section V.

\section{Preliminaries}
For a signal $s \in {L^2}(R)$ and even window $g \in {L^2}(R)$, the STFT is defined as:
\begin{equation}
V(t,\omega ) = \int\limits_{ - \infty }^\infty  {s(\tau )g(\tau  - t){e^{ - j\omega (\tau  - t)}}d\tau }.
\end{equation}
The signal $s(t)$ can be reconstructed by
\begin{equation}
s(t){\rm{ = }}\frac{1}{{2\pi g(0)}}\int\limits_{ - \infty }^\infty  {V(t,\omega )} d\omega.
\end{equation}
If there exists a sufficiently small number $\varepsilon $, satisfying $\left| {A'(t)} \right| < \varepsilon$ and $\left| {\varphi ''(t)} \right| < \varepsilon $, then the STFT of signal $s(t) = A(t){e^{j\varphi (t)}}$ can be expressed as
\begin{equation}
V(t,\omega ){\rm{ = }}A(t){e^{j\varphi (t)}}\hat g(\omega  - \varphi '(t))
\end{equation}
where $\hat g(\omega)$ is the Fourier transform of $g(t)$.
Then, the spectrum of STFT is obtained by
\begin{equation}
\left| {V(t,\omega )} \right|{\rm{ = }}A(t)\hat g(\omega  - \varphi '(t)).
\end{equation}
The spectrum is symmetric about the IF $\varphi '(t)$ and its coefficients are clustered around $\varphi '(t)$.
From formula (4), it is clearly seen that the performance of the STFT depends on $\hat g(\omega )$. 

The ideal TFR of signal $s(t) = A(t){e^{j\varphi (t)}}$ that we desire to achieve is
\begin{equation}
{\rm{ITFA}}(t,\omega ){\rm{ = }}A(t){e^{j\varphi (t)}}\delta (\omega  - \varphi '(t))
\end{equation}
However, it is unrealizable to achieve the ideal TFR by STFT. It requires $g(t) = 1$, in contradiction to the conditions of the window.

To approximate the ideal TFR and preserve the ability to reconstruct the signal, the SST was proposed, which is defined as
\begin{equation}
Ts(t,\xi ) = \int\limits_{ - \infty }^\infty  V (t,\omega )\delta (\xi  - \omega (t,\omega ))d\omega 
\end{equation}
where $\omega (t,\omega )$ is the estimated IF.

The original signal $s(t)$ can be recovered by
\begin{equation}
s(t){\rm{ = }}\frac{1}{{2\pi g(0)}}\int\limits_{\left| {\xi  - \omega (t,\omega )} \right| \le \Gamma } {Ts(t,\xi )d\xi }
\end{equation}

\section{PROPOSED METHOD}
In \cite{daubechies2011synchrosqueezed}, the adaptive TF decomposition is modeled as a variational problem in which one seeks to minimize
\begin{multline}
\mathop {\min }\limits_{F(t,\omega )} \int {{{\left| {\operatorname{Re} \left[ {\int {F(t,\omega )d\omega } } \right] - s(t)} \right|}^2}dt} \\
 + \lambda {\iint {\left| {{\partial _t}F(t,\omega ) - i\omega F(t,\omega )} \right|}^2}dtd\omega 
\end{multline}
where $F(t,\omega ) \in {L^2}({R^2})$.
The literature \cite{daubechies2011synchrosqueezed} states that the solution of variational problem is the result of SST, i. e., $F(t,\omega ){\rm{ = }}T(t,\xi )$, letting $\omega {\rm{ = }}\xi $.

The SST performance based on STFT is limited by the estimated IF
\begin{equation}
\omega (t,\omega ) = \frac{{\frac{\partial }{{\partial t}}V(t,\omega )}}{{jV(t,\omega )}}.
\end{equation}
However, the results of formula (9) do not work well, leading to smearing on the TF plane. In many studies, $V(t,\omega )$ is replaced by other better TFA tools to improve SST performance, but we believe that it is not the most effective method.

To enhance the quality of the TFR, the variational model in formal (8) improved as
\begin{multline}
\mathop {\min }\limits_{F(t,\omega )} \underbrace {\int {{{\left| {\operatorname{Re} \left[ {\int {F(t,\omega )d\omega } } \right] - s(t)} \right|}^2}dt} }_{\text{I}} \\
+ \underbrace {\lambda \iint {\left| {F(t,\omega ) - {\text{ITFA}}(t,\omega )} \right|}dtd\omega. }_{{\text{II}}}
\end{multline}
The aim of this model is to find $F(t,\omega )$ such that it not only approximates the ideal TFR but also allows the reconstruction of signal. 
The role of part I of this model is to maintain the reconstruction capability, and part II serves to approximate the ideal TFR.

However, the model in (10) is hard to solve, so additional restrictions need to be added to simplify the model.

\textbf{Simplification 1:} For a TFR, if the reassignment only occurs on the frequency axis, the reassignment result still retains the ability to recover signal.
Therefore, if the reassignment is restricted to the frequency direction, the model in (10) is simplified to
\begin{equation}
\mathop {\min }\limits_{F(t,\omega )} \underbrace {\iint {\left| {F(t,\omega ) - {\text{ITFA}}(t,\omega )} \right|}dtd\omega. }_{{\text{II}}}
\end{equation}

Then, we consider a multicomponent signal $s(t) = \sum\limits_{k = 1}^K {{s_k}(t)} {\text{ = }}\sum\limits_{k = 1}^K {{A_k}(t){e^{j{\varphi _k}(t)}}}$, whose ideal TFR is
\begin{equation}
{\text{ITFA}}(t,\omega ){\text{ = }}\sum\limits_{k = 1}^K {{A_k}(t){e^{j{\varphi _k}(t)}}\delta (\omega  - {{\varphi '}_k}(t))}. 
\end{equation}
The STFT of signal $s(t)$ is
\begin{equation}
V(t,\omega ){\text{ = }}\sum\limits_{k = 1}^K {{A_k}(t){e^{j{\varphi _k}(t)}}\hat g(\omega  - {{\varphi '}_k}(t))}. 
\end{equation}

From formulas (12) and (13), it is clear that to approximate the ideal TFR using the results of STFT,  it is only necessary to reallocate the coefficients of STFT to the IF ${{{\varphi '}_k}(t)}={\omega _k}(t,\omega )$ of signal $s_k(t)$.

Assuming that the estimation of ${\varphi '_k}(t)$ is ${\hat \omega _k}(t,\omega )$, the novel reassignment operator is defined as
\begin{equation}
T(t,\xi ) = \sum\limits_{k = 1}^K  \int\limits_{ - \infty }^\infty  V (t,\omega )\delta (\xi  - {\hat \omega _k}(t,\omega ))d\omega 
\end{equation}

Substituting (13) into (14), we can obtain
\begin{multline}
\begin{gathered}
 \int\limits_{ - \infty }^\infty  {\sum\limits_{k = 1}^K {{A_k}(t){e^{j{\varphi _k}(t)}}\hat g(\omega  - {{\varphi '}_k}(t))} \delta (\xi  - {{\hat \omega }_k}(t,\omega ))d\omega }  \hfill \\
{\text{ = }}\sum\limits_{k = 1}^K {\int\limits_{ - \infty }^\infty  {{A_k}(t){e^{j{\varphi _k}(t)}}\hat g(\omega  - {{\varphi '}_k}(t))\delta (\xi  - {{\hat \omega }_k}(t,\omega ))d\omega } }  \hfill \\
{\text{ = }}\sum\limits_{k = 1}^K {{A_k}(t){e^{j{\varphi _k}(t)}}\delta (\xi  - {{\hat \omega }_k}(t,\omega ))\int\limits_{ - \infty }^\infty  {\hat g(\omega  - {{\varphi '}_k}(t))d\omega } }  \hfill \\
=2\pi g(0)  \sum\limits_{k = 1}^K {{A_k}(t){e^{j{\varphi _k}(t)}}\delta (\xi  - {{\hat \omega }_k}(t,\omega ))}  \hfill \\ 
\end{gathered}
\end{multline}
where $\int\limits_{ - \infty }^\infty  {\hat g(\omega  - {{\varphi '}_k}(t))d\omega }=2\pi g(0) $. This conclusion is easy to prove and we will not discuss it here.

Therefore, (14) can be reformulated as
\begin{equation}
T(t,\xi ) =2\pi g(0)\sum\limits_{k = 1}^K {{A_k}(t){e^{j{\varphi _k}(t)}}\delta (\xi  - {{\hat \omega }_k}(t,\omega ))}
\end{equation}
In comparison with formula (12) and formula (16), one can find that the novel reassignment operator achieves the ideal TFR.

Further, we will prove that the new reassignment operator is invertible.
\begin{multline}
\begin{aligned}
&\int_{ - \infty }^\infty  {2\pi g(0)\sum\limits_{k = 1}^K {{A_k}(t){e^{j{\varphi _k}(t)}}\delta (\xi  - {{\hat \omega }_k}(t,\omega ))} d\xi } \\ 
&=2\pi g(0) \sum\limits_{k = 1}^K {{A_k}(t){e^{j{\varphi _k}(t)}}\int_{ - \infty }^\infty  {\delta (\xi  - {{\hat \omega }_k}(t,\omega ))d\xi } }\\
&=2\pi g(0) \sum\limits_{k = 1}^K {{A_k}(t){e^{j{\varphi _k}(t)}}} \\ 
&=2\pi g(0) s(t)\\
\end{aligned}
\end{multline}

So, the signal can be reconstructed by
\begin{equation}
s(t){\text{ = }}\frac{1}{{2\pi g(0)}}\int\limits_{ - \infty }^\infty  {T(t,\xi )d\xi } 
\end{equation}

From formulas (12) and (16), one can know that to make $T(t,\xi )$ approximate ${\text{ITFA}}(t,\omega )$, only ${\hat \omega _k}(t,\omega )$ needs to approximate ${\varphi '_k}(t)$. Thus, we can reduce the model in (11) again.

\textbf{Simplification 2:}
Setting $\omega {\text{ = }}\xi $, the model in (11) can be simplified again as
\begin{equation}
\min \sum\limits_{k = 1}^K {\iint {\left| {{{\hat \omega }_k}(t,\omega ) - {\omega _k}(t,\omega )} \right|dtd\omega }}
\end{equation}
Therefore, the performance of $T(t,\xi )$ depends on the result of IF estimation ${\hat \omega _k}(t,\omega )$. This also determines that $T(t,\xi )$ can eliminate the energy diffusion on the TF plane. Therefore, the key of our proposed method is to estimate IF. There have been many advanced methods for IF estimation, such as ridge extraction algorithms \cite{yu2019local}. 

In essence, our method is to squeeze the TF coefficients to IF, rather than to the vicinity of IF like SST. Fig. 1 clearly shows the manner of allocating coefficients for several post-processing tools. The SET extracts only the TF coefficients at estimated IF, and LMSST squeezes the coefficients to their maximum values in certain frequency intervals.

\begin{figure}[h!]
	\centering
	\subfloat[SST]{
		\includegraphics[width=0.22\textwidth]{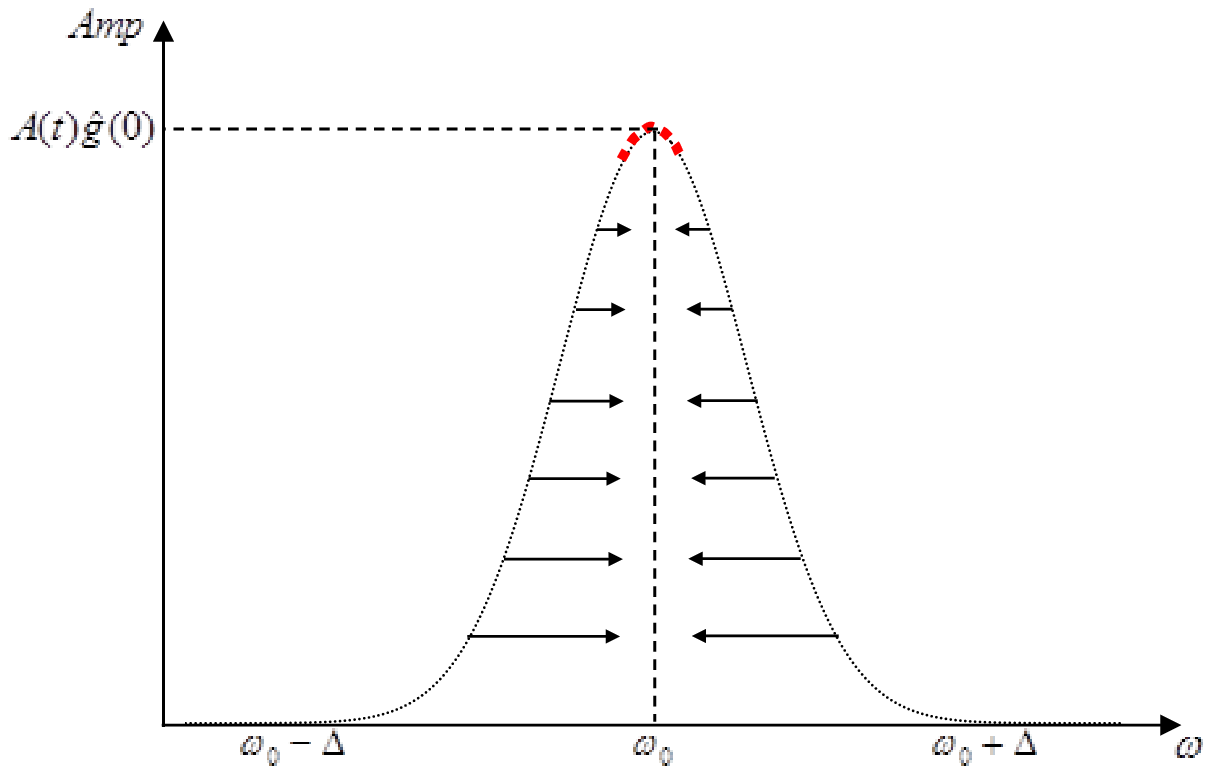}
		%\caption{fig1}%6.25
	}
	\subfloat[SET]{
		\includegraphics[width=0.22\textwidth]{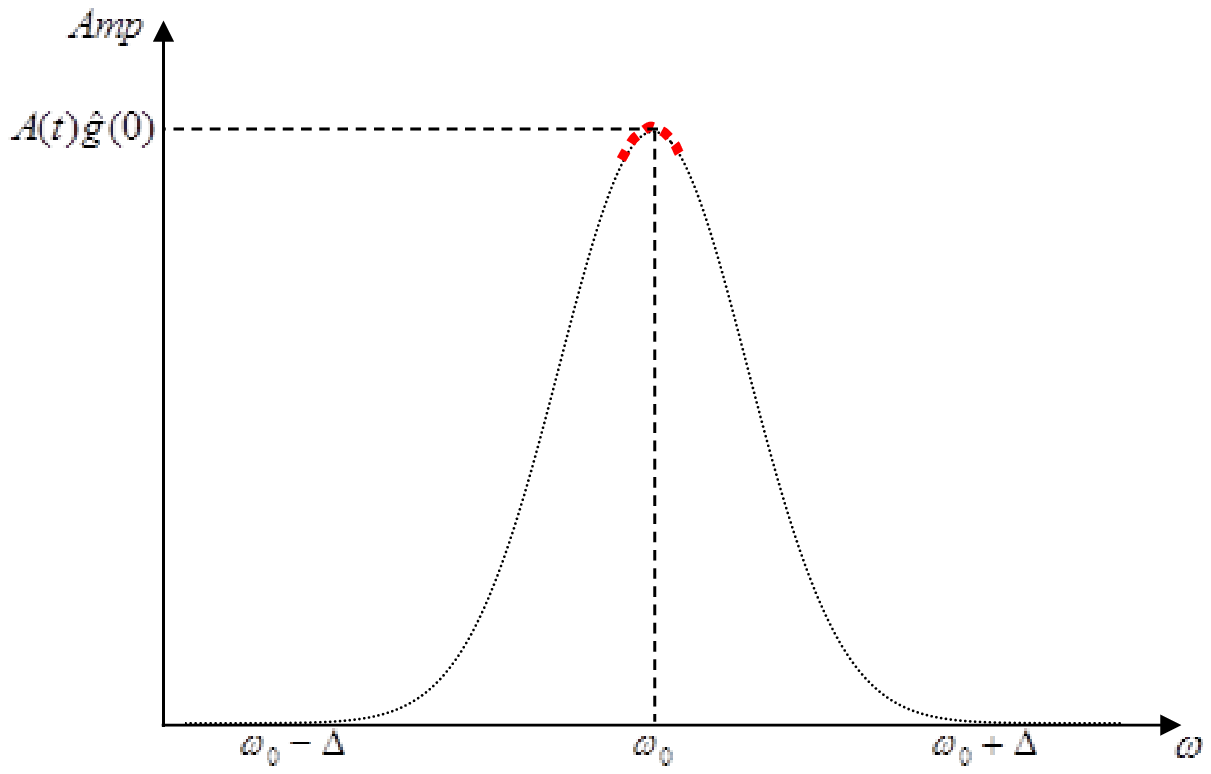}
		%\caption{fig1}%6.25
	}
	\quad
	\subfloat[LMSST]{
		\includegraphics[width=0.22\textwidth]{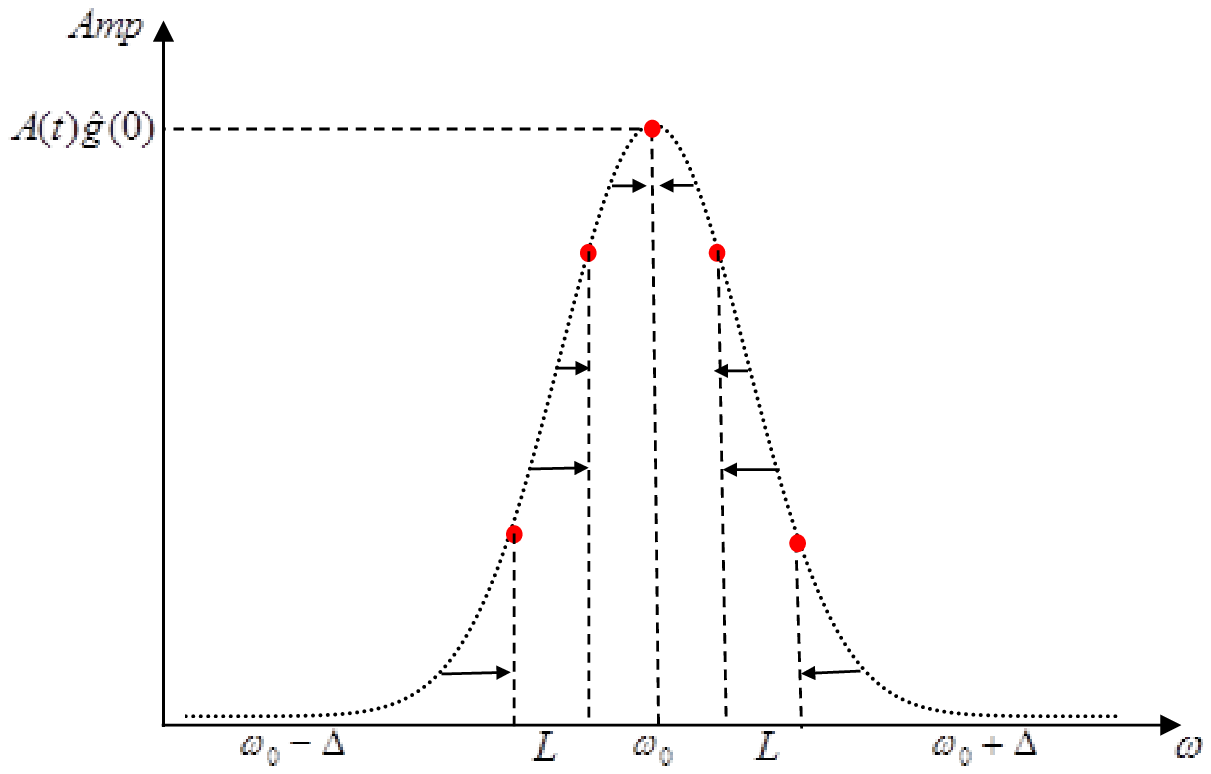}
		%\caption{fig1}%6.25
	}
	\subfloat[Ours]{
		\includegraphics[width=0.22\textwidth]{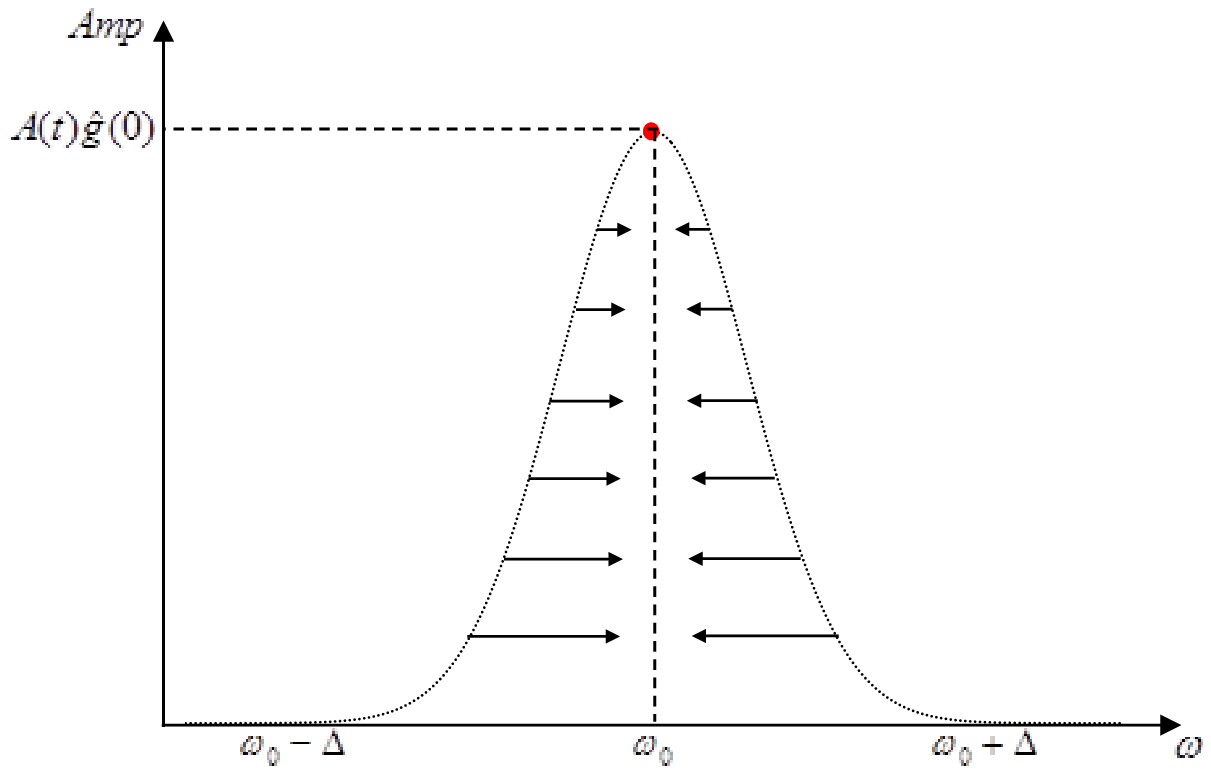}
		%\caption{fig1}%6.25
	}
	\caption{The manner of allocating coefficients for several post-processing tools.}
\end{figure}

Finally, we give the modular and high-resolution TF post-processing technique.
For an arbitrary TFR $G(t,\omega )$, assume that the signal is reconstructed by
\begin{equation}
s(t) = \rho  \int\limits_{ - \infty }^\infty  {G(t,\omega )d\omega } 
\end{equation}
where $\rho$ is a reconstruction factor, which can be a number or function. It depends on the specific TFR used. For example, $\rho$ is $\frac{1}{{2\pi g(0)}}$ in STFT.

The post-processing technique based on $G(t,\omega )$  is defined as:
\begin{equation}
{T_G}(t,\xi ) = \sum\limits_{k = 1}^K  \int\limits_{ - \infty }^\infty  G (t,\omega )\delta (\xi  - {\hat \omega _k}(t,\omega ))d\omega 
\end{equation}
The reconstruction of the original signal from ${T_G}(t,\xi )$ can be accomplished by
\begin{equation}
s(t) = \rho \int\limits_{ - \infty }^\infty  {{T_G}(t,\xi )d\xi }.  
\end{equation}

Our method is called modular for two reasons. (1) The STFT $V(t,\xi )$ can be replaced by any TFA tool $G(t,\omega )$, such as the short-time fractional Fourier transform, adaptive STFT, S-transform and chirplet transform, etc. (2) The estimated IF ${\hat \omega _k}(t,\omega )$ can be replaced by any one of the estimated methods, which directly affects performance of ${T_G}(t,\xi )$. The combination of $G(t,\omega )$ and ${\hat \omega _k}(t,\omega )$ with each other will produce many interesting results, but there is no difference in essence.

In this letter, a simple and effective estimation approach is applied, which has the advantage of not requiring any priori information about the number of signal components. The local maxima of the TFR $G(t,\omega )$ are taken as an estimate of IF, i.e.
\begin{equation}
\left\{ {{{\hat \omega }_1}(t,\omega ), \cdots ,{{\hat \omega }_M}(t,\omega )} \right\} = \mathop {{\text{arglmax}}}\limits_\omega  \left| {G(t,\omega )} \right|
\end{equation}
where $\mathop {{\text{arglmax}}}\limits_{\omega}  \left| {G(t,\omega )} \right|$ represents a set of local maximum points. However, the number of estimated IF by formula (23) may exceed the number of real signal components due to interference. Therefore, we can perform a simple filter on $G(t,\omega )$ to solve this problem.
\begin{equation}
{G(t,\omega )} {\text{ = }}\left\{ \begin{gathered}
{G(t,\omega )} ,\left| {G(t,\omega )} \right| > \gamma \max (\left| {G(t,\omega )} \right|) \hfill \\
0,other \hfill \\ 
\end{gathered}  \right.
\end{equation}
where $\gamma$ serves as a filter.
\section{NUMERICAL VALIDATIONS}
In this section, our proposed method is compared with STFT and other post-processing tools, such as RM, SST, SET and LMSST, for both synthetic and real world data.
\subsection{The Multicomponent Signal with Distinct FM-AM}
A multicomponent signal with distinct FM-AM is defined as
\begin{multline}
\begin{aligned}
&{s_1}(t) = \sin (2\pi (40t + sin (4\pi t))\\
&{s_2}(t) = \sin (2\pi (10t + 10{(t - 0.5)^3})
\end{aligned}
\end{multline}
whose sampling frequency is 128 Hz and time of 1 s.
\begin{figure}[htbp]
	\centering
		\subfloat[STFT]{
		\includegraphics[width=0.22\textwidth]{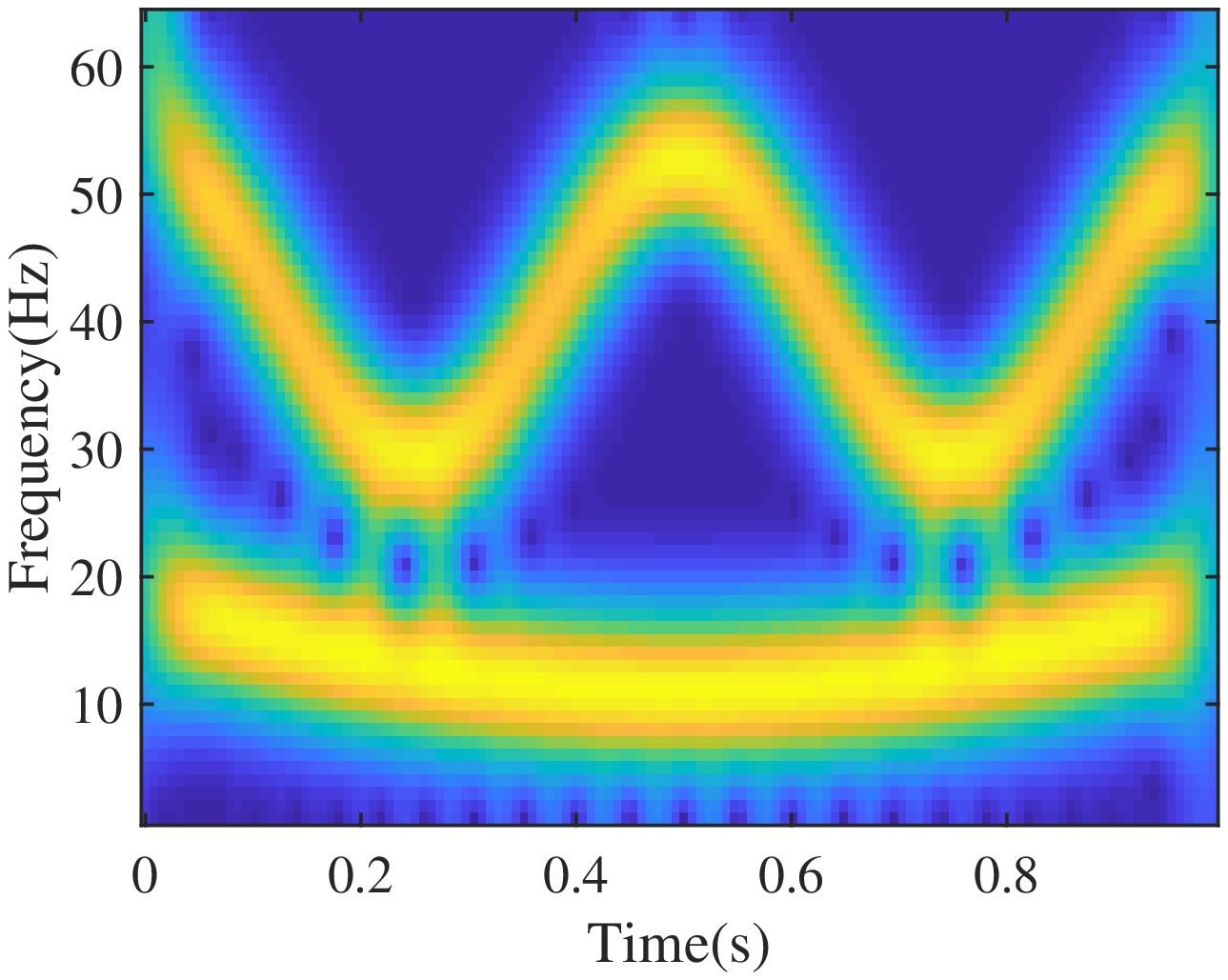}
		%\caption{fig1}%6.25
	}
	\subfloat[SST]{
		\includegraphics[width=0.22\textwidth]{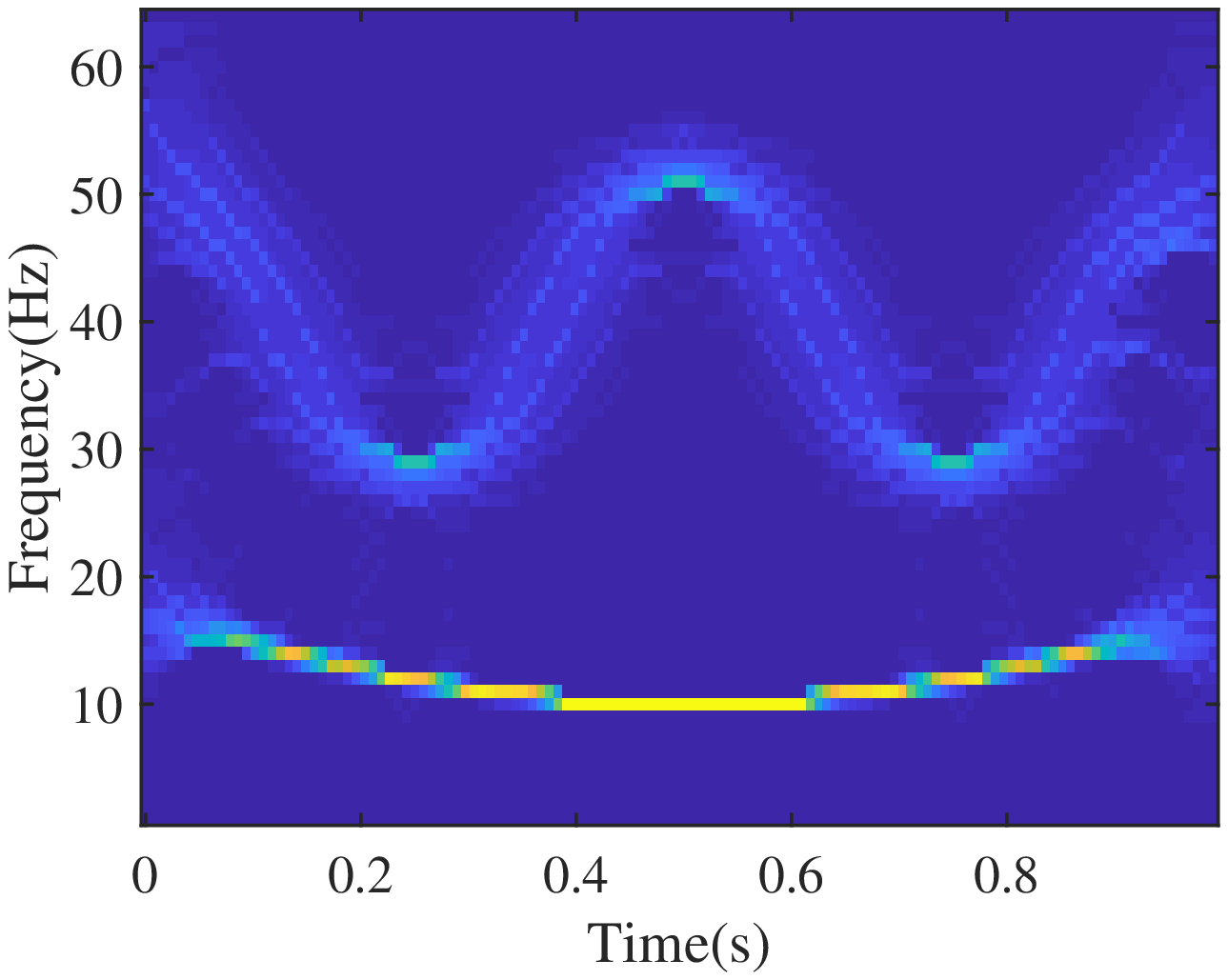}
		%\caption{fig1}%6.25
	}
\quad
	\subfloat[RM]{
	\includegraphics[width=0.22\textwidth]{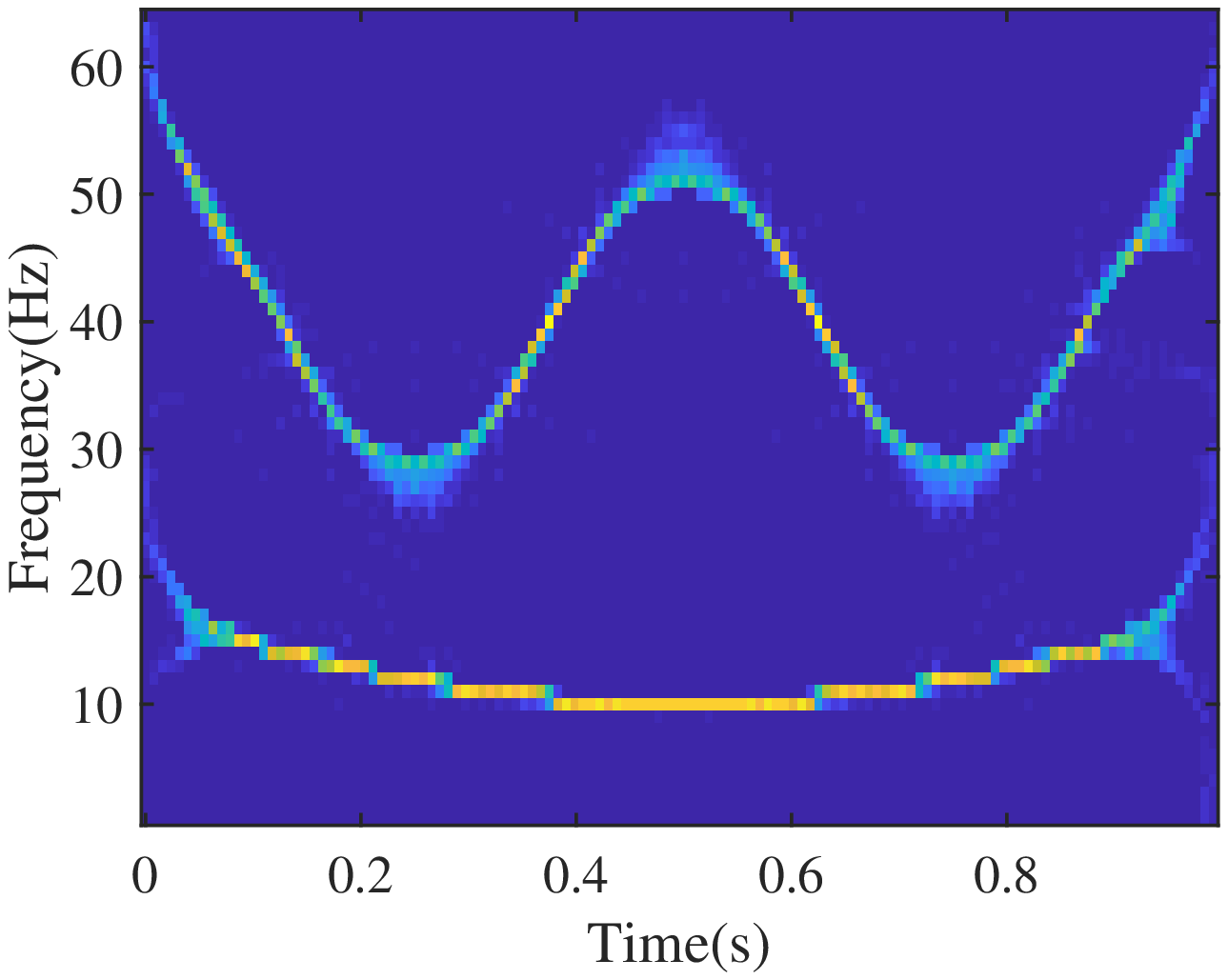}
	%\caption{fig1}%6.25
	}
	\subfloat[SET]{
		\includegraphics[width=0.22\textwidth]{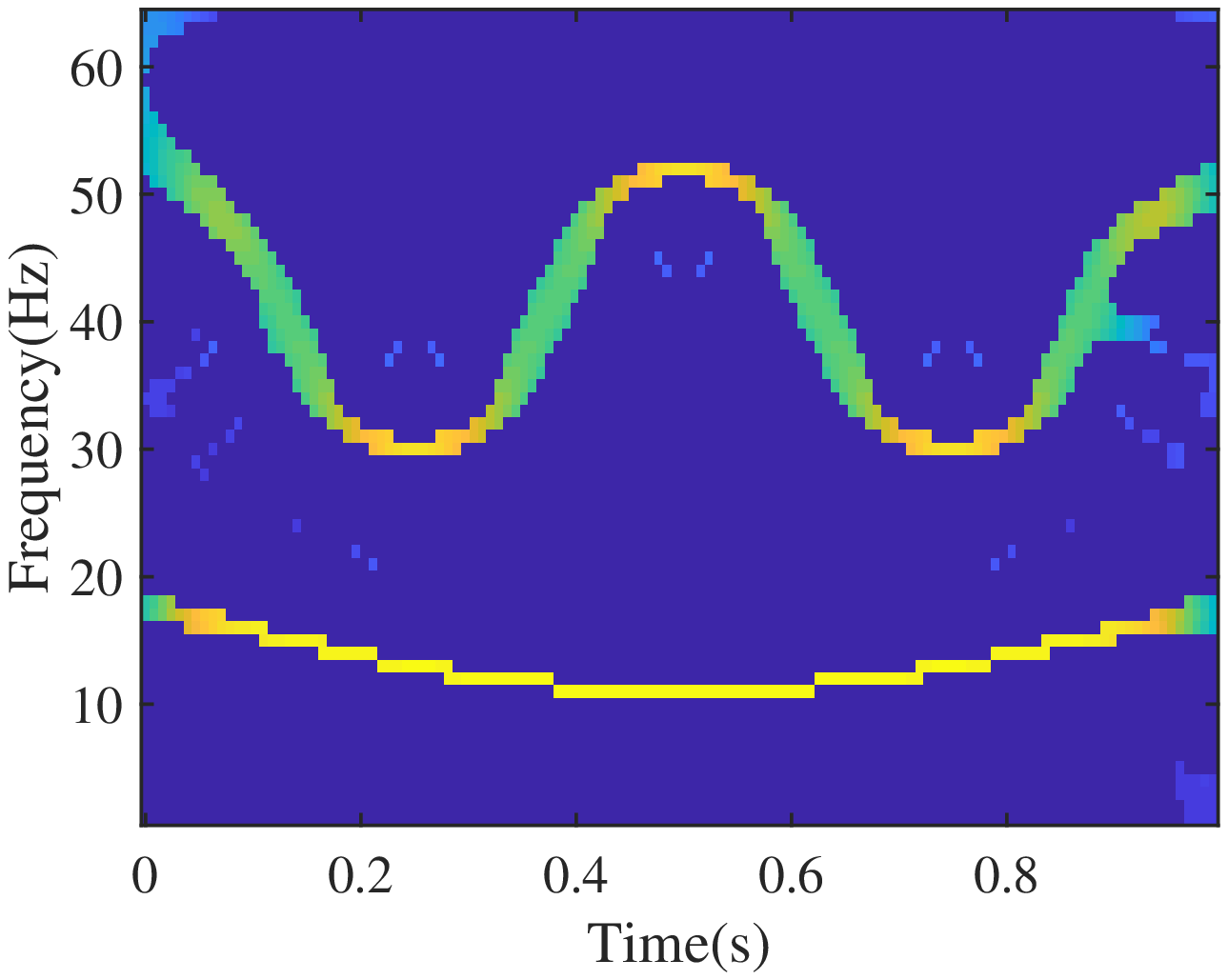}
		%\caption{fig1}%6.25
	}
	\quad
	\subfloat[LMSST]{
		\includegraphics[width=0.22\textwidth]{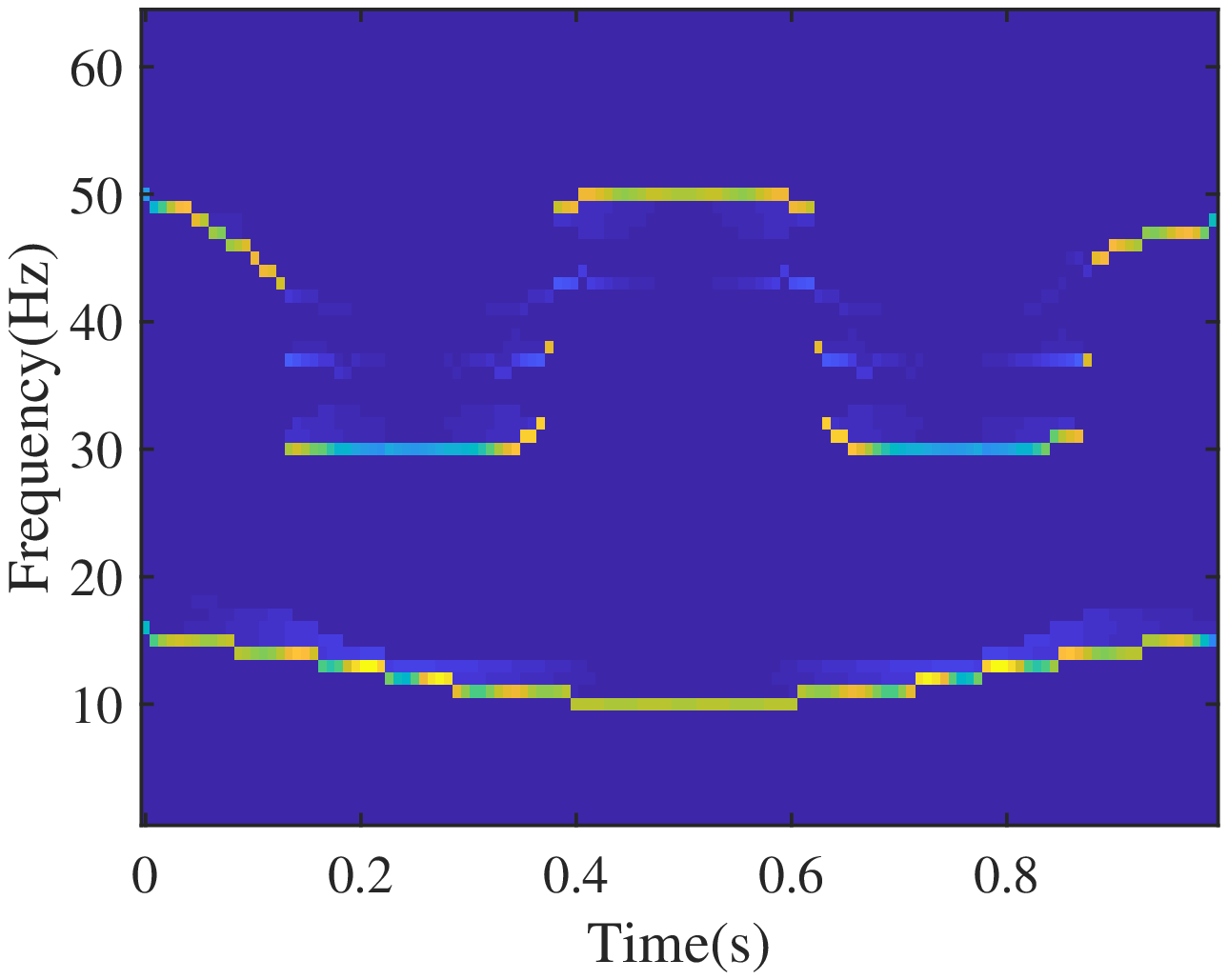}
		%\caption{fig1}%6.25
	}
	\subfloat[Ours]{
		\includegraphics[width=0.22\textwidth]{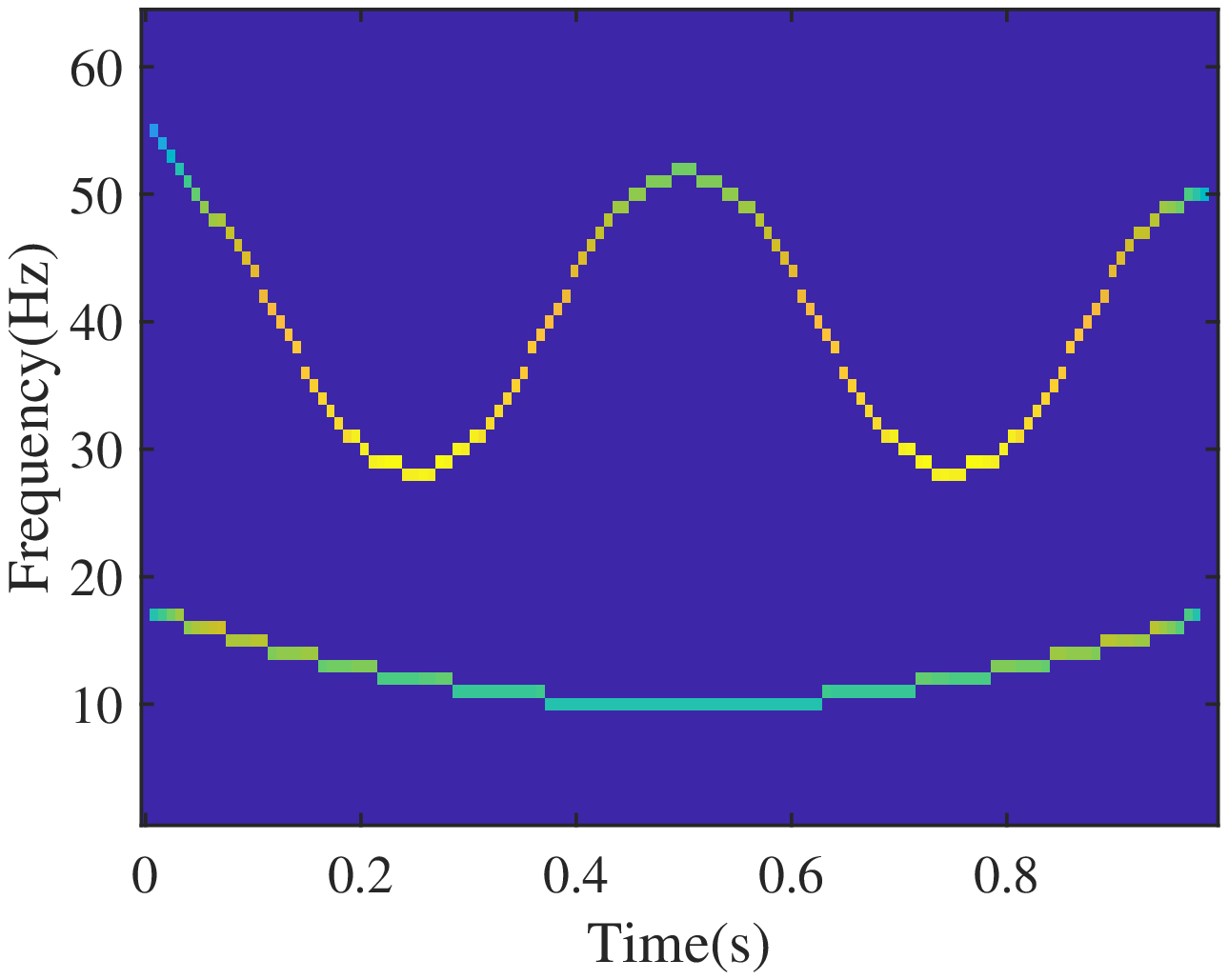}
		%\caption{fig1}%6.25
	}
	%\caption{The experimental results of multicomponent signal were obtained by (a) STFT, (b) SST (c) RM, (d) SET, (e) LMSST and (f) Our method.}
	\caption{The experimental results of multicomponent signal.}
\end{figure}

The results of TFR are shown in Fig. 2, which shows that our method achieves the most ideal result.The result of STFT suffers from poor TF resolution and the results of SST and SET occur with severe smearing. The result  of LMSST is distorted where the frequency changes rapidly. Although RM is relatively effective compared to other post-processing method, it is irreversible.

%%\begin{table}[htbp]
%	\renewcommand{\arraystretch}{1.5}
%	\centering
%	\caption{The R\'{e}nyi entropy of multicomponent signal with distinct FM-AM}
%	\label{tab2.1}
%	\begin{tabular}{ccccccc}%l=left, r=right,c=center分别代表左对齐，右对齐和居中，字母的个数代表列数
%		\hline
%		TFA & STFT\qquad&SST \qquad& RM\qquad  & SET& LMSST\qquad  & ModSST \\
%		\hline
%		R\'{e}nyi entropy & 12.1647 \qquad &9.2885\qquad & 9.3149 \qquad &8.7728& 8.1904 \qquad &7.9602 \qquad\\
%		\hline
%	\end{tabular}
%\end{table}

\subsection{The Multicomponent Signal with Crossover IF}
In this section, we consider a multicomponent signal with crossover IF to verify the effectiveness of our method. The signal is modeled as
\begin{multline}
\begin{aligned}
&{s_1}(t) = {e^{j500\pi t}} \hfill \\
&{s_2}(t) = {e^{ - 0.5t}}{e^{j(2\pi (500\pi t + 100\sin (2\pi t))}} \hfill \\
&{s_3}(t) = 0.8{e^{0.5t}}{e^{j(2\pi (500\pi t - 100\sin (2\pi t))}} \hfill \\ 
\end{aligned}
\end{multline}
whose sampling frequency is $2^{10}$ Hz and the signal duration is 1 s.

To show the characteristics of modularity, we used the IF estimation method described in \cite{chen2017separation} and \cite{chen2017intrinsic}. Our method can effectively solve the IF crossover problem, which is attributed to the excellent IF estimation. Furthermore, the TF results of our proposed method can be viewed as the trajectory of IF. The relationship between IF estimation and TF rearrangement can also be seen here. It can be concluded that the rearrangement results can be used to estimate IF, and the estimated IF can be utilized to reallocate the TF coefficients. Thus, formula (23) can be replaced by alternative methods of IF estimation to suit some sophisticated tasks.
\begin{figure}[htbp]
	\centering
	\subfloat[STFT]{
		\includegraphics[width=0.22\textwidth]{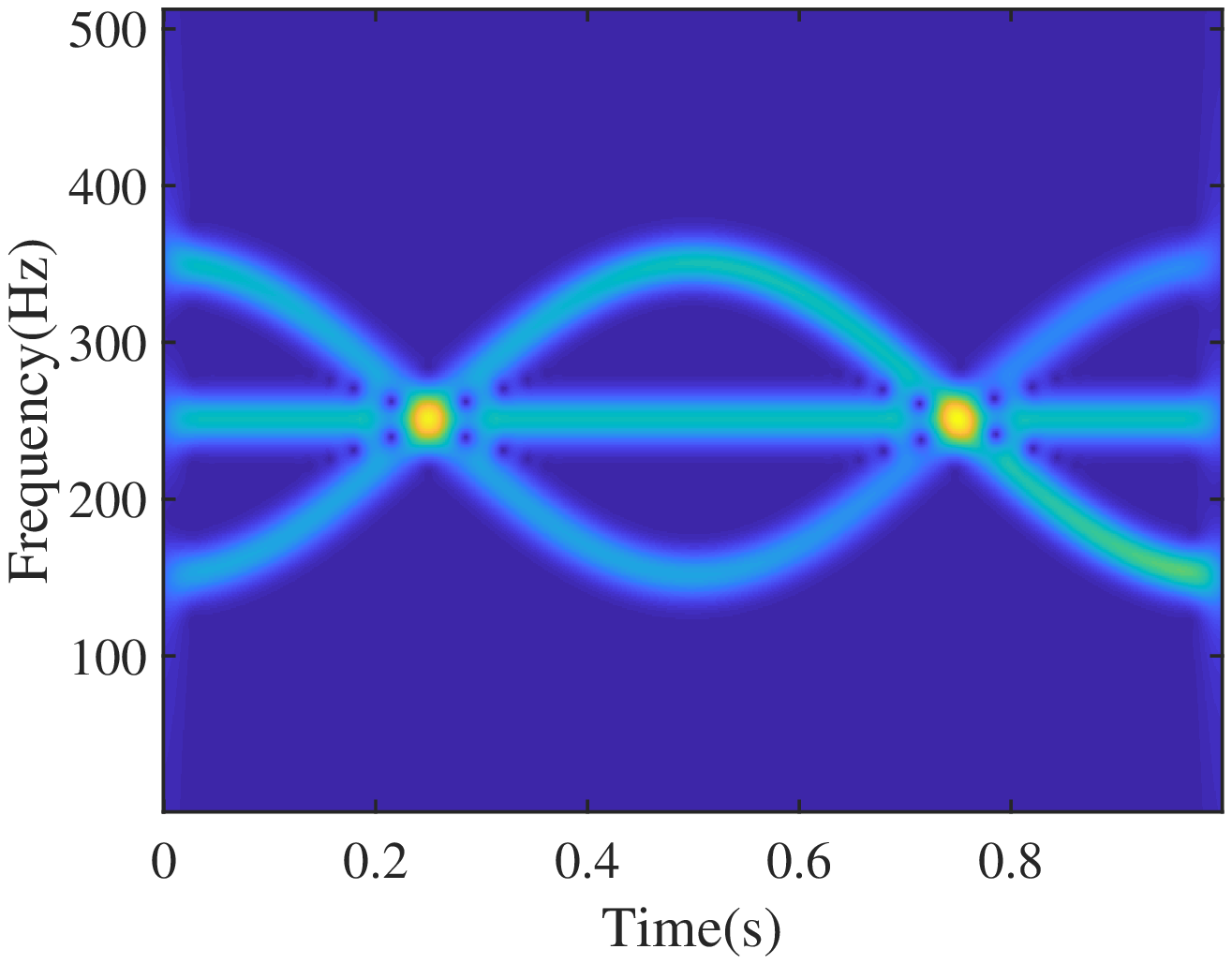}
		%\caption{fig1}%6.25
	}
	\subfloat[SST]{
		\includegraphics[width=0.22\textwidth]{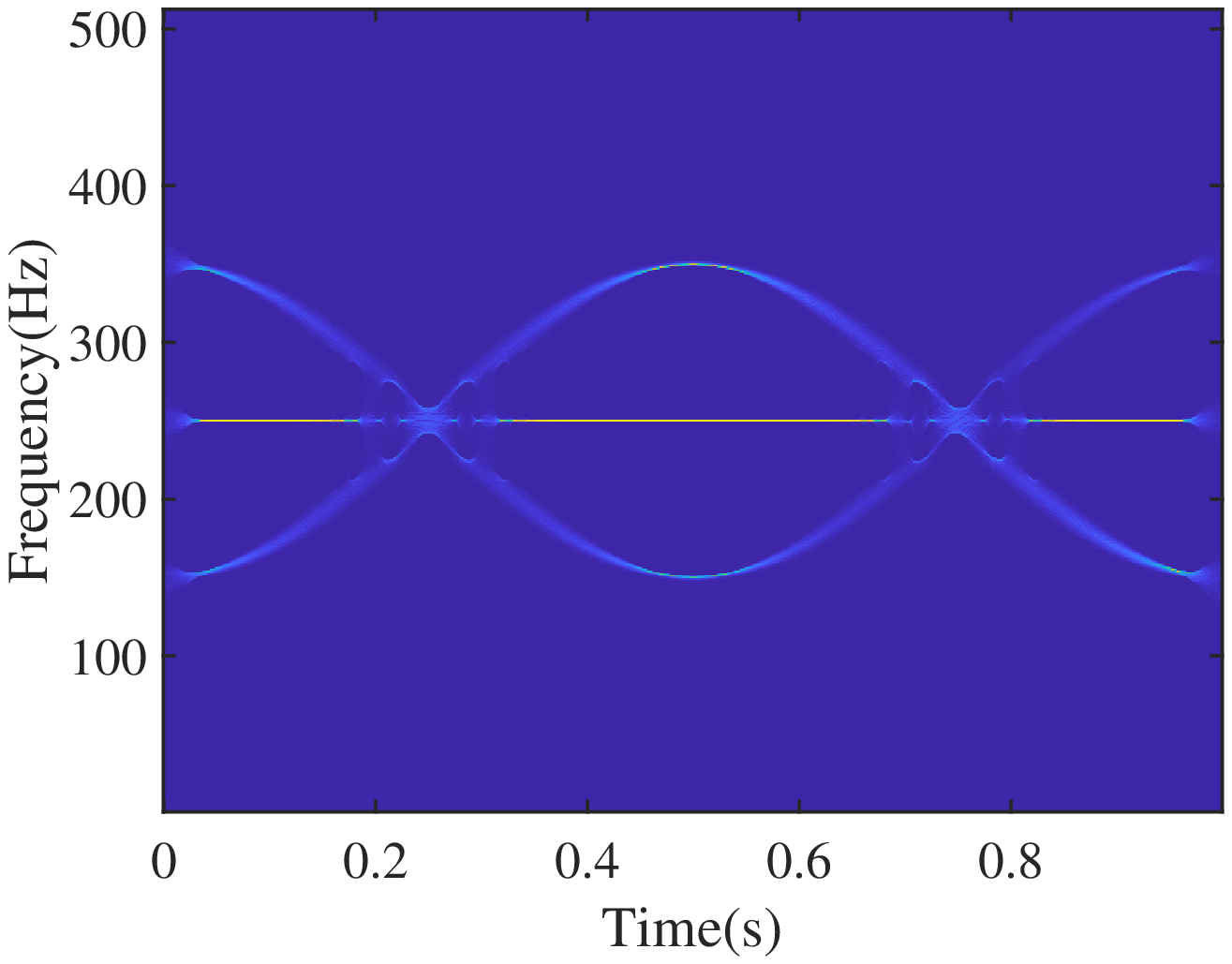}
		%\caption{fig1}%6.25
	}
	\quad
	\subfloat[RM]{
		\includegraphics[width=0.22\textwidth]{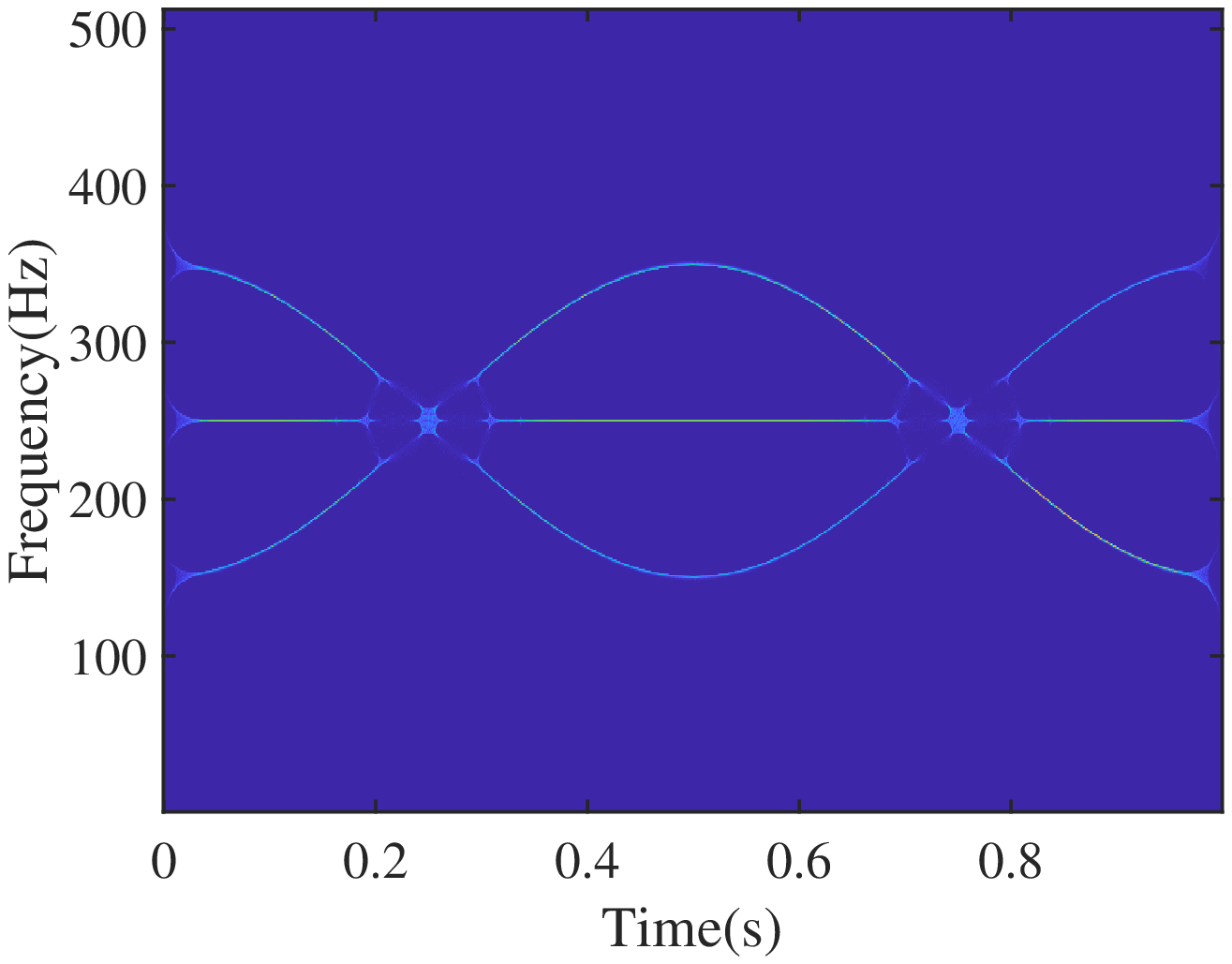}
		%\caption{fig1}%6.25
	}
	\subfloat[SET]{
		\includegraphics[width=0.22\textwidth]{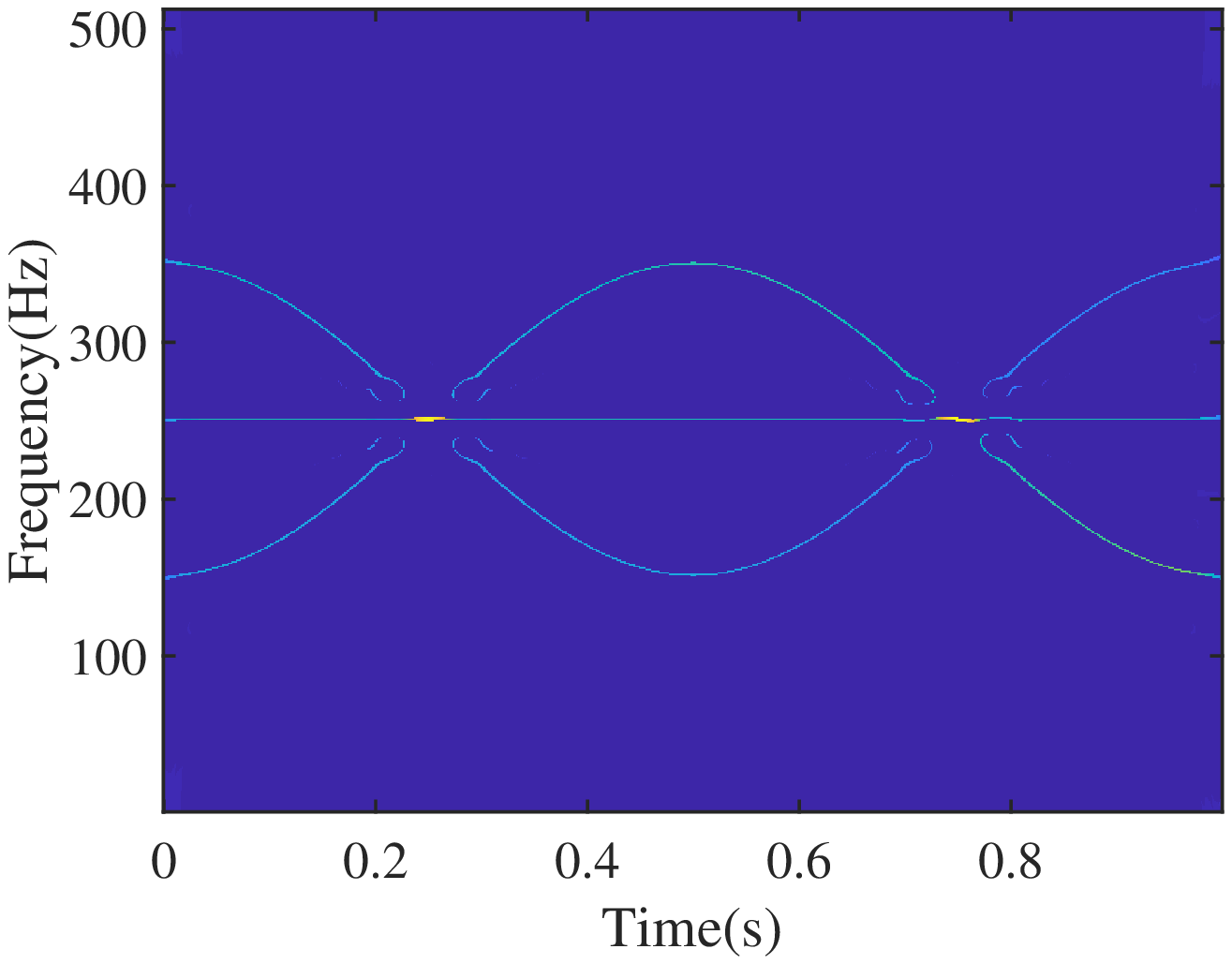}
		%\caption{fig1}%6.25
	}
	\quad
	\subfloat[LMSST]{
		\includegraphics[width=0.22\textwidth]{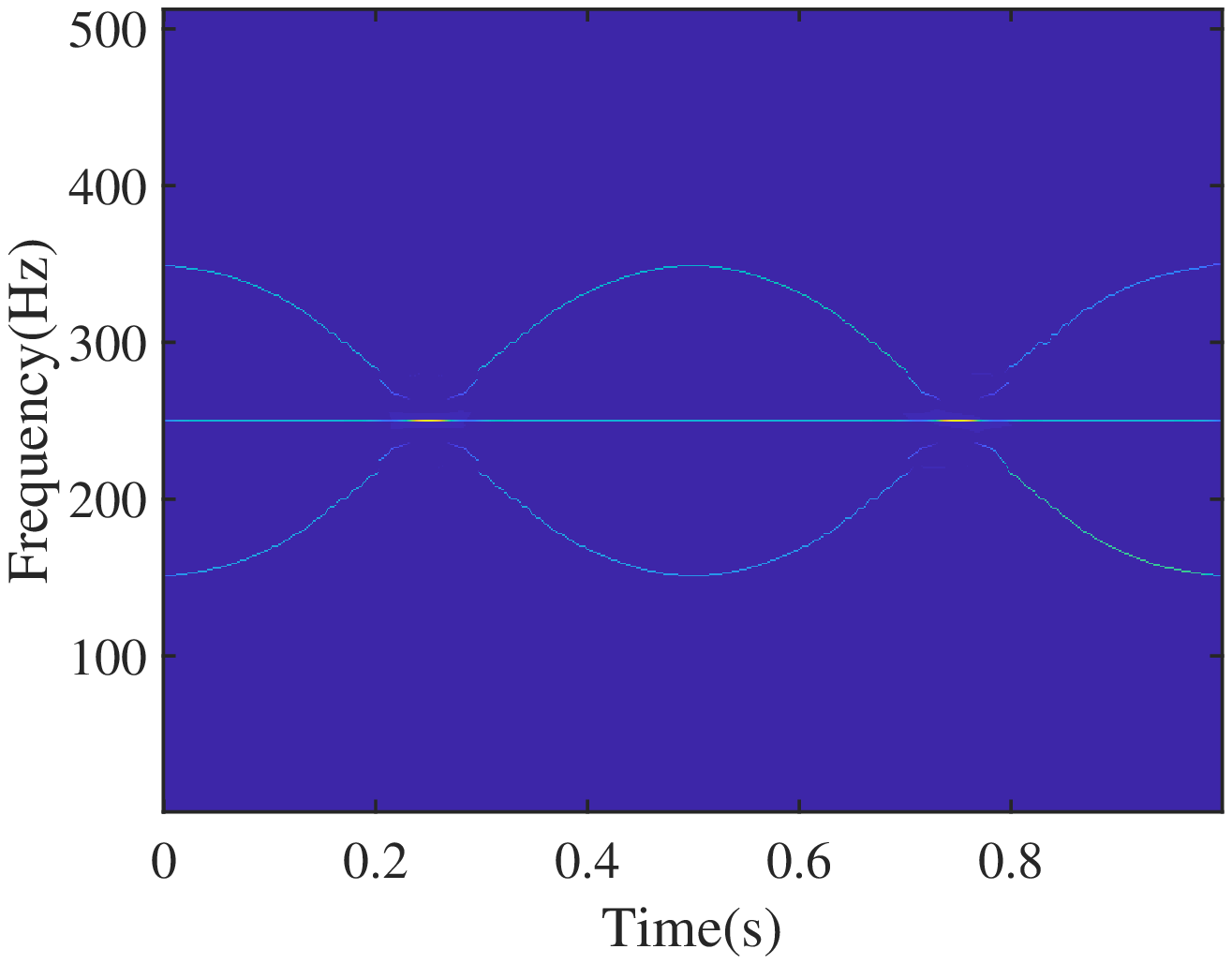}
		%\caption{fig1}%6.25
	}
	\subfloat[Ours]{
		\includegraphics[width=0.22\textwidth]{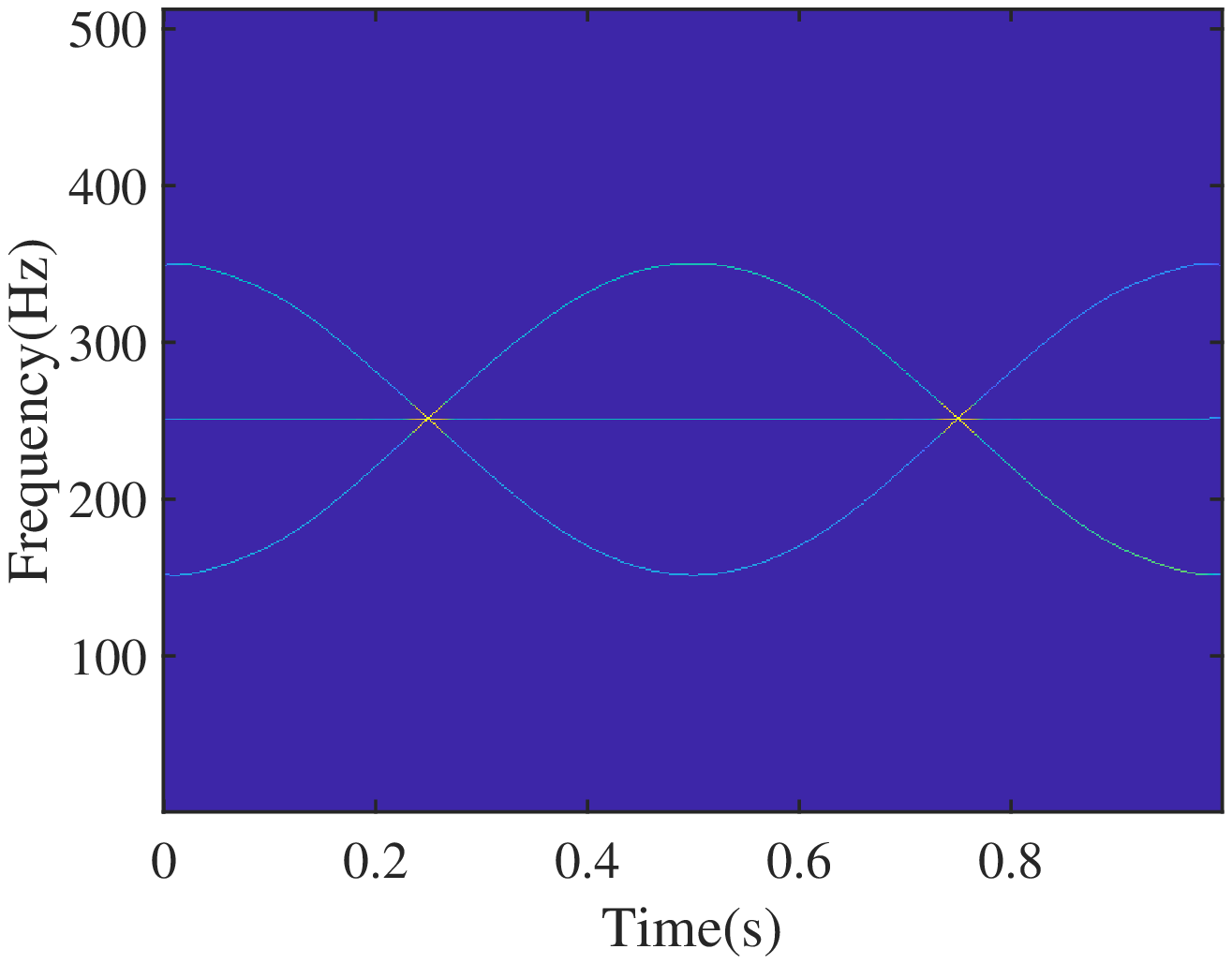}
		%\caption{fig1}%6.25
	}
%	\caption{The experimental results of signal with Crossover IF were obtained by (a) STFT, (b) SST (c) RM, (d) SET, (e) LMSST and (f) Our method.}
	\caption{The experimental results of signal with Crossover IF.}
\end{figure}

%\begin{table}[htbp]
%	\renewcommand{\arraystretch}{1.5}
%	\centering
%	\caption{The R\'{e}nyi entropy of multicomponent signal with Crossover IF}
%	\label{tab2.2}
%	\begin{tabular}{ccccccc}%l=left, r=right,c=center分别代表左对齐，右对齐和居中，字母的个数代表列数
%		\hline
%		TFA & STFT\qquad&SST \qquad& RM\qquad  & SET& LMSST\qquad  & ModSST \\
%		\hline
%		R\'{e}nyi entropy & 16.6054 \qquad &12.5777\qquad & 11.8739 \qquad &12.0283& 11.8628 \qquad &11.3906 \qquad\\
%		\hline
%	\end{tabular}
%\end{table}

\subsection{Gravitational-Wave Signal}
To further validate the performance of our method, we apply it to gravitational-wave signal, whose specific description can be found in \cite{pham2017high}. The results of TFR are shown in Fig. 4.

Compared with other methods, our method and LMSST can obtain sharper TF results. However, some small frequency changes are easy to be ignored by LMSST. For an unknown signal, we believe any change is cause for concern. Our method can capture any small changes on the TF plane. For noisy signals, the disturbance can be filtered by adjusting the value of $\gamma $ in (24) to reduce the impact of noise. Additionally, the result of LMSST produced artifacts in interval $t \in [0.43,0.45]$ with rapid shift of IF. Therefore, our method has potential to be applied to many fields of interest.
\begin{figure}[h!]
	\centering
	\subfloat[STFT]{
		\includegraphics[width=0.23\textwidth]{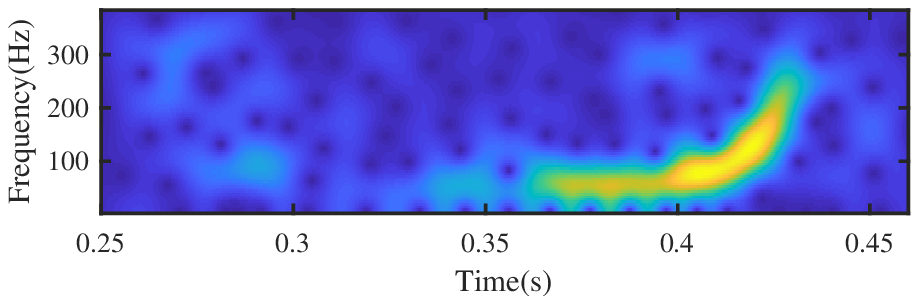}
		%\caption{fig1}%6.25
	}
	\subfloat[SST]{
		\includegraphics[width=0.23\textwidth]{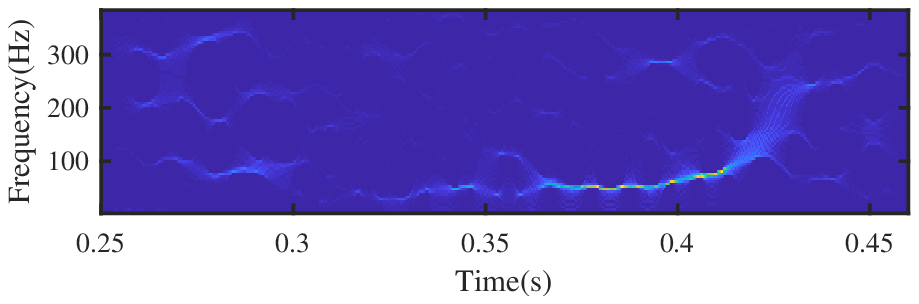}
		%\caption{fig1}%6.25
	}
\quad
	\subfloat[RM]{
		\includegraphics[width=0.23\textwidth]{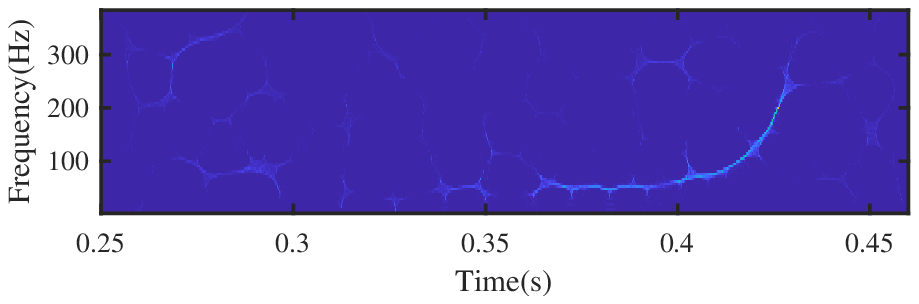}
		%\caption{fig1}%6.25
	}
	\subfloat[SET]{
		\includegraphics[width=0.23\textwidth]{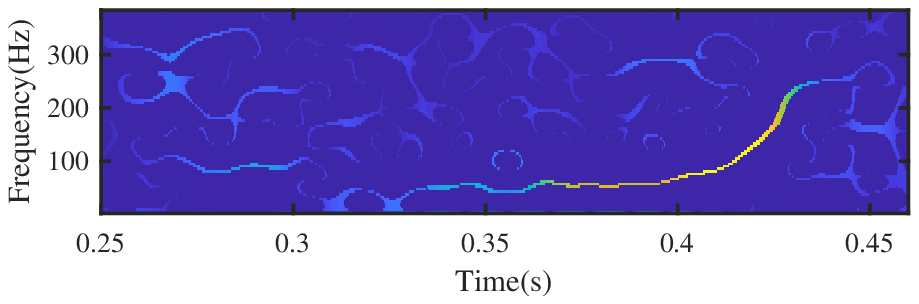}
		%\caption{fig1}%6.25
	}
\quad
	\subfloat[LMSST]{
		\includegraphics[width=0.23\textwidth]{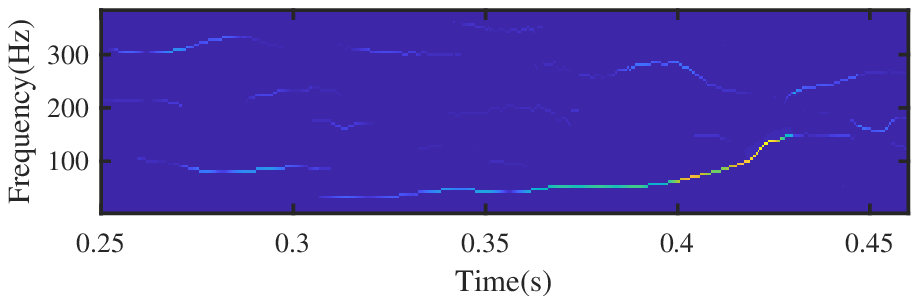}
		%\caption{fig1}%6.25
	}
	\subfloat[Ours]{
		\includegraphics[width=0.23\textwidth]{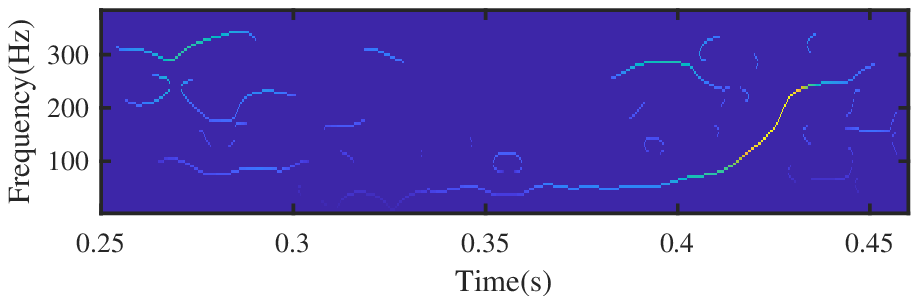}
		%\caption{fig1}%6.25
	}
	%\caption{The experimental results of gravitational-wave signal were obtained by (a) STFT, (b) SST (c) RM, (d) SET, (e) LMSST and (f) Our method.}
	\caption{The experimental results of gravitational-wave signal.}
\end{figure}

%\begin{table}[h!]
%	\renewcommand{\arraystretch}{1.5}
%	\centering
%	\caption{The R\'{e}nyi entropy of gravitational-wave signal}
%	\label{tab2.3}
%	\begin{tabular}{ccccccc}%l=left, r=right,c=center分别代表左对齐，右对齐和居中，字母的个数代表列数
%		\hline
%		TFA & STFT\qquad&SST \qquad& RM\qquad  & SET& LMSST\qquad  & ModSST \\
%		\hline
%		R\'{e}nyi entropy & 16.6330 \qquad &13.8811\qquad & 13.2067\qquad &12.9834& 11.8752 \qquad &11.7208 \qquad\\
%		\hline
%	\end{tabular}
%\end{table}

\section{Conclusion}
In this letter, a new TFA technique is proposed to achieve the ideal TFR. We have established a variational model to obtain an adaptive TF decomposition. After simplifying the model, we have derived the relationship between IF estimation and ideal TFR, and also clarified the relationship between IF estimation and reassignment. The future work should be to expand our technique to the time-scale plane. It would also be interesting to use a better IF estimation algorithm instead of the method proposed in this letter to improve the quality of the TFR.

\bibliographystyle{ieeetr}
\IEEEtriggeratref{11} % 表示对第35个参考文献开始换栏
\bibliographystyle{IEEEtran}
\bibliography{document}

\end{document}